\documentclass[%
 reprint,
superscriptaddress,
 amsmath,amssymb,
 aps,
]{revtex4-1}
\usepackage[colorlinks,linkcolor=blue, citecolor=blue, urlcolor=blue]{hyperref}%pour colorer liste figures, ref...ET CLIQUER DESSUS dans le pdf
\usepackage{graphicx}% Include figure files
\usepackage{dcolumn}% Align table columns on decimal point
\usepackage{bm}% bold math
\usepackage{subfigure}
\newcommand\vect[1]{\bm{#1}}
\newcommand\dd{\mathrm{d}}

\newcommand\rr{\bm{r}}
\newcommand\LL{\mathcal{L}}
\newcommand\refeq[1]{Eq. \eqref{#1}}
\newcommand\reffig[1]{Fig. \ref{#1}}
\newcommand\jaco{\text{sn}}
\begin{document}

\preprint{APS/123-QED}

%\title{A mixed phase field-atomistic multi-scale approach to predict patterned microstructures in irradiated materials: application to AgCu}% 

\title{Prediction of irradiation induced microstructures in the AgCu model alloy via atomistic and phase field modelling}% 

\author{G. Demange}
 \email{gilles.demange@univ-rouen.fr}
 \affiliation{DEN/DMN/SRMA/LA2M, CEA Saclay, 91191 Gif-sur-Yvette, France}%Lines break automatically or can be forced with \\
\author{L. Lun\'eville}
% \email{laurence.luneville@cea.fr}
\affiliation{DEN/SERMA/LLPR, CEA Saclay, 91191 Gif sur Yvette, France}%
\author{V. Pontikis}%
% \email{vassilis.pontikis@cea.fr}
\affiliation{ DEN/DMN/LSI, CEA Saclay, 91191 Gif-sur-Yvette, France}
\author{D. Simeone}%
% \email{david.simeone@cea.fr}
\affiliation{DEN/DMN/SRMA/LA2M, CEA Saclay, 91191 Gif-sur-Yvette, France}
\date{\today}

\begin{abstract}

In this work, a multiscale approach based on phase field was developped to simulate the microstructure's evolution under irradiation in binary systems, from atomic to microstructural scale. For that purpose, an efficient numerical scheme was developed. In the case of AgCu alloy under Krypton ions irradiation, phenomenological parameters were computed using atomistic methods, as a function of the temperature and the irradiation flux. As a result, we predicted the influence of the irradiation flux and the temperature on the formation of patterned microstructures. In the case of AgCu, our model was qualitatively and quantitatively validated by a diffraction experimental study.

\end{abstract}

\pacs{Valid PACS appear here}% PACS, the Physics and Astronomy
\maketitle

%\tableofcontents

\section{Introduction}
\label{sec:intro}

Steady state patterned microstructures have lately sparked interest amidst the scientist community \cite{simeone2006}. Such microstructures consist in predominance domains of bounded characteristic dimension. These domains can display a wide variety of morphologies, from labyrinthine patterns in polymer mixtures \cite{Harada}, to quasi-spherical Ni$_3$Al precipitates under ion irradiation \cite{schmitz2001phase}. The latter microstructure results from a ballistic disordering induced by the slowing down of impinging ions \cite{cross1993}. Recently, ion beam mixing process emerged as a proper candidate to manufacture patterned microstructures \cite{bernas2003}, for   microelectronic industry notably \cite{cheng1990}. The advantage of this fabrication process is that the microstructure features can be controlled by such parameters as irradiation temperature and ion flux. 

However, a quantitative description of the irradiation induced microstructure remains extremely challenging for two reasons. First, irradiated materials are brought very far from equilibrium, and their evolution cannot be described within the standard framework of statistical thermodynamics. Second, several time and space scales are involved in the process: starting from ballistic displacements induced by irradiation ($10^{-12}$ s, $1-10$ \AA{}), passing by diffusion mechanisms ($10^{-6}$), up to precipitates growth (over years at millimetric scale). This is why, so far, this issue remains out of reach for atomistic approaches alone, such as Ab Initio methods and molecular dynamics \cite{Stoller2000,Krasnochtchekov2005}. 

The present study aims at achieving greater understanding of materials microstructure far from equilibrium, where few analytical results can be found, at a relevant time and space scale. In the specific case of irradiated materials, it intends to classify the material microstructures by the mean of a phase diagram in the temperature-irradiation flux space, and more specifically the temperature and flux leading to patterned microstructures in binary alloys. 

This can only be achieved provided the number of freedom degrees are reduced, and space and time scales are crossed. The phase field method \cite{landau1950,cahn1961,hohenberg1977} meets these requirements. This approach has notably proved itself efficient to study aging in binary alloys \cite{lecoq2011numerical}. Yet, it did not take irradiation effects into account in the first place. To overcome this limitation, Martin asserted that the thermal spike in a cascade could be taken into account by an athermal source term \cite{martin1984}. This idea was then invested by Bellon and al. \cite{Averback2010}. Using a spectral approach, they could predict the formation of precipitates of stationary size due to ballistic effects, depending on reduced irradiation parameters \cite{bellontheo}. Alternatively, a self-consistent method provided with greater accuracy in  \cite{champ}. Yet, these studies were performed in adimensionalized coordinates, and no clear link to the temperature and the irradiation flux could be established, albeit Enrique and al. made a first attempt to overpass this limitation in \cite{Enrique2003}. Using a mixed molecular dynamics-kinetic Monte Carlo approach, they were able to demonstrate the possibility of disrupted coarsening for realistic temperatures and flux. However, their study remained also qualitative. Actually,  so far, no complete scale jump, from atomic to microstructural scale could be performed, in order to simulate the formation of patterned microstructure under irradiation.  In particular, no computation of the  domain in the temperature-flux plane was achieved.

In this work, the phase field method was developed to simulate the microstructure's evolution under irradiation in binary systems, at the relevant time and space scales. As a result, a new phase field model taking irradiation effects into account was set and physically justified. An efficient numerical scheme was developed to simulate the long stage microstructure in reduced units. Atomistic methods were developed to compute the phenomenological parameters, as a function of temperature and irradiation flux, for the AgCu model alloy under Krypton ions irradiation. The different microstructures of ion irradiated alloys were classified in a temperature-flux phase diagram. This diagram was eventually qualitatively and quantitatively validated by a diffraction experimental study \cite{Wei1997}. The choice of the AgCu system was motivated by the presence of a miscibility gap. Besides, this system comes with numerous experimental data on the thermodynamical properties of its components and the alloy, as well as various numerical studies to compare with, notably in \cite{Enrique2003}. This makes AgCu a convenient application system. In this case, the microstructure consists in silver precipitates in a copper matrix.

This work is structured as follows. First, the phase field model, where a ballistic term and dependence of atomic mobility on irradiation flux was added, is presented. Second, the numerical methods at use are detailed. Then, results are displayed in the following order. The phenomenological parameters in the phase field model are estimated for AgCu under Krypton ions irradiation. Then, the phase diagram  of the irradiated system provided by our numerical phase field approach in reduced coordinates is displayed. Finally, this diagram is expressed in the flux-temperature plane, and comparison with a diffraction study from the literature is discussed.

%%%%%%%%%%%%%%%%%%%%%%%%%%%%%%%%%%%%%%%%%%%%%%%%%%%%%%%%%%%%%%%%%%%%%%%%%%%%%%%%%%%%%%%%%%%%%%%%%%%%%%%%%%%%%%%%%%%
\section{Phase field model with a ballistic contribution}
\label{sec:model}

\subsubsection{Equilibrium model}
\label{subsubsec:bal}

Within the phase field model \cite{chen2002phase}, the microstructure of binary alloys can be characterized by the coarse grained concentration $c(\rr,t)$ of one component. In the case of phase separation, the evolution of the concentration can be modelled using the Model B equation \cite{hohenberg1977}: 

\begin{equation}
\label{ch}
\frac{\partial c}{\partial t} = \bar{M}\nabla^2\left[\frac{\delta F}{\delta c}(c)\right],
\end{equation}
where $\bar{M}$ is the atomic mobility, and $F(c)$ is the free energy. Equation \refeq{ch} describes the system evolution from a non equilibrium state to equilibrium, through minimisation of its free energy, via diffusional mechanisms. For a heterogeneous system, $F$ can be written \cite{landau1950}:

\begin{equation}
\label{nrjtot}
F(c)=\int_{\Omega} \left[f_h(c)+\frac{\kappa}{2}|\nabla c|^2\right]\dd \Omega.
\end{equation}
Here, $f_h(c)$ is the homogeneous free energy density, and  $\kappa|\nabla c|^2/2$  is the energy cost of diffuse interface formation in an heterogeneous medium. $\kappa>0$ is the stiffness parameter, which is proportional to the interface energy. Based on Lifshitz's criterion for the small fluctuations of the concentration from their concentration in liquid state $c_L$, $f_h$  can be expanded in Taylor series with respect to the concentration deviation $c-c_L$:

\begin{equation}
\label{nrjhomo}
f_h(c)=\frac{a_2(T)}{2}(c-c_L)^2+\frac{a_3(T)}{3} (c-c_L)^3+\frac{a_4(T)}{4}(c-c_L)^4.
\end{equation}
Contrary to the standard Cahn-Hilliard equation \cite{cahn1961}, a cubic term was added to $f_h$, so that first order phase transitions could be accounted for \cite{toledano1996}. In particular, this term will be required in the case of AgCu. $f_h$ displays two minima corresponding to two stable equilibrium states. These minima are defined by Hazewinkel's approach for stationary solutions of \refeq{ch} in the case of non symmetric potentials \cite{hazewinkel1986pattern}. At equilibrium, $c_+$ and $c_-$ satisfy $f_h'(c_+)=f_h'(c_-)$, $f_h''(c_+)=f_h''(c_-)$ and $f_h''(c_{\pm})>0$, where function $f_h$ is derived with respect to the concentration. The last condition ensures the stability of both phases at equilibrium. This provides with:

\begin{equation}
\label{solub}
c_{\pm}=-\frac{a_3}{3a_4}\pm \sqrt{\frac{1}{3}\left(\frac{a_3}{a_4}\right)^2-\frac{a_2}{a_4}}+c_L.
\end{equation}
In \refeq{ch}, the atomic mobility $\bar{M}$ plays a dominant part in the definition of the time scale of the dynamics. It was determined in Onsager's formalism \cite{allnatt2003}, under Darken's approximation \cite{darken1948}. As is, this led $\bar{M}$ to  depend strongly on $c$, through the factor $c(1-c)$ (see  appendix \ref{ap:mob}). This factor actually restrains diffusion to interfaces only, as $c(1-c)\to 0$ far from interfaces. This corresponds to interface diffusion. It is characterized by a $t^{1/4}$ growth law of precipitates \cite{PhysRevB.52.R685}. However, experiments observed a $t^{1/3}$ growth law \cite{Ges1997},  associated to bulk diffusion. The $c(1-c)$ factor was hence averaged, so that Martin's standard expression for $\bar{M}$ \cite{martin1990} was recovered: 

\begin{equation}
\label{mobth}
\bar{M}=\frac{\bar{c}(1-\bar{c})}{k_B T} \tilde{D}.
\end{equation}
Here, $\bar{c}$ is the mean concentration, and $\tilde{D}$ is the chemical interdiffusion coefficient \cite{allnatt2003}.  $\tilde{D}$ theoretically depends on $c(\rr,t)$ as well. Yet, in the case of similar isotopic diffusion coefficients of both species in the alloy \cite{butrymowicz1976}, it can be approximated by an average isotopic diffusion coefficient, thus independent of $c$ \cite{demangethese2015}. This assumption is satisfied by the AgCu alloy, due to very close creation and migration energies \cite{Foiles1986,Nordlund1998,Adams1988,Wynblatt1969,McGervey1975}. 

\subsubsection{Irradiation effects}
\label{subsubsec:bal}

In our study, we made the assumption that irradiation impacted the microstructure of an alloy through two main mechanisms: ballistic effects induced by atomic relocations, and diffusion enhancement due to point defects creation.

As for ballistic effects, Martin suggested that they could be described by an athermal term \cite{martin1984}. The local fluctuations of concentration induced by irradiation could then be expressed by a local matter balance \cite{simeone2010}:  

\begin{equation}
\label{bal}
\frac{\partial c}{\partial t}=\Gamma \times (p_R*c-c),
\end{equation}
where $\Gamma$ is an atomic relocation frequency, and $p_R$ a probability distribution of displacements in subcascades. It involves Sigmund and Gras Marti ion mixing formalism  \cite{sigmund1981}. Ion mixing asserts that ballistic effects can be treated as an homogeneous source term. The justification relates to the strong gap of time scale between the ballistic stage of a displacements cascade ($\sim 100$ ps), and precipitates growth driven by diffusion ($\sim 1$ ms). The evolution of the microstructure under irradiation is thus the result of two mecanisms, acting separately on two different time scales. This assumption requires that the covering volume fraction of the cascade by subcascades, $f_r$, exceeds the percolation threshold \cite{Misaelides1994}:

\begin{equation}
\label{percol}
f_r= n_{\text{sc}}\times \frac{l_{\text{sc}}}{L_{\text{c}}}\geq 0.2.
\end{equation}
Here, $n_{\text{sc}}$, $l_{\text{sc}}$ and $L_{\text{c}}$ are the number of subcascades within the cascade, the width of the cascade and its length respectively.  This makes the microstructure  only sensible to the average effect of numerous cascades covering one another. The strong heterogeneity due to subcascades \cite{Antoshchenkova2015168} can hence be passed over, and every point $\rr$ in the homogeneous cascade can be considered as the center of a subcascade. Therefore, $p_R$ does not depend on the position, but on the flight distance $r$ between the initial and final position of the displaced atom only. We assumed that $p_R$ followed an exponential decay, mimicking small displacements in subcascades:

\begin{equation}
\label{loidep1}
p_R(r)=\frac{1}{2\pi R^2} \exp\left(\frac{-r}{R}\right).
\end{equation}
Here $R$ is the mean relocation distance. As for $\Gamma$, it is the product of the irradiation flux $\Phi$, and a relocation macroscopic cross section $\sigma^d$:

\begin{equation}
\label{gamma}
\Gamma=\sigma^d \Phi, \quad \sigma^d=\frac{V_{\text{cell}}}{N_{\text{cell}}}\times \frac{N_d}{ L_{\text{c}}}.
\end{equation}
Here, $V_{\text{cell}}$ and $N_{\text{cell}}$ are the cell volume and number of atoms per cell. $N_d$ is the number of relocated atoms per cascade.

The second phenomenon related to irradiation is the enhancement of atomic diffusion, due to point defects recreation. Introducing additional vacancies and interstitials concentration $\delta c_V^{\text{irr}}$ and $\delta c_{I}^{\text{irr}}$, the irradiation contribution $M^{\text{irr}}$ to the mobility can be written \cite{sizmann1978}:

\begin{equation}
\label{mobirr}
M^{\text{irr}}= \frac{\bar{c}(1-\bar{c})}{k_B T} \left[\delta c_V^{\text{irr}}D_V+\delta c_{I}^{\text{irr}}D_I \right],
\end{equation}
where $D_V$ and $D_I$ are the vacancy and interstitial diffusion coefficients in the alloy. This expression requires that $D_V$ and $D_I$ are very close for each species of the alloy \cite{demangethese2015}. This assumption is one again satisfied in the case of AgCu system. The values of $\delta c_V^{\text{irr}}$ and $\delta c_{I}^{\text{irr}}$ were estimated by the chemical model of Sizmann \cite{sizmann1978}. Thermal recombination, and sink absorption at interfaces were also taken into account \cite{barbu}. As point defects migration takes place on a very small time scale ($\sim 10^{-9}$ s) in comparison with precipitates growth ($\sim 10^{-6}$ s), their concentration can be set at their equilibrium value. It is defined by the final state of Sizmann's equations (see appendix \ref{ap:mob}). The total mobility $M$ then reads:

\begin{equation}
\label{mobtot}
M=\bar{M}+M^{\text{irr}}.
\end{equation}
In this case, $M$ depends on both temperature and irradiation flux. Its expression is given in appendix \ref{ap:mob}. Finally introducing ballistic term of \refeq{bal} in \refeq{ch}, the Cahn-Hilliard equation under irradiation becomes:

\begin{equation}
\label{chtot}
\frac{\partial c}{\partial t} = M\nabla^2\left[f_h'(c) -\kappa\nabla^2 c\right]+\Gamma (p_R*c-c).
\end{equation}
Let us remark that in the phase field model, atomic fluctuations were passed over by the coarse graining procedure. Strictly speaking, they should be accounted for by the \textit{ad-hoc} introduction of a noise, consistently with Cook's approach \cite{Cook1970}. However, this study focuses on low temperatures, where noise effects upon the microstructure are small. They were hence neglected in this work.

\subsubsection{Adimensionalized equation and space and time scales}

To model the microstructural evolution in an irradiated material, the phase field equations were rewritten in their reduced form first. The reduced concentration $\phi$ is defined from the solubility limits $c_{\pm}$ displayed in \refeq{solub}, so as to vary between $-1$ and $+1$ in the absence of  irradiation and without asymmetry in the free energy:

\begin{equation}
\label{minimahomo}
\phi=\frac{c-c_L}{\alpha},  \quad \text{where} \quad \alpha=\frac{c_+ - c_-}{2}.
\end{equation}
Using \refeq{chtot}, the evolution of $\phi$ is given by the equation:

\begin{equation}
\label{chphi}
\frac{\partial \phi}{\partial t'}=\nabla^2\left[[3\lambda^2-1]\phi-3\lambda \phi^2+\phi^3-\nabla^2\phi\right]+W (p_{R'}*\phi-\phi).
\end{equation}
This procedure reduces the number of parameters from 6 ($a_2$, $a_3$, $a_4$, $\Gamma$, $R$, $M$) to 3 ($\lambda$, $W$ and $R'$), without loss of information. These 3 reduced parameters are linked to the 6 original ones, through the phase field time and space scales $t_0$ and $l_0$: 

\begin{equation}
\label{adim1}
\left\{
\begin{aligned}
& t_0=\frac{|a_2|\xi^2}{M a_4^2\alpha^4 V_{\text{at}}}, \quad l_0=\frac{1}{\alpha}\sqrt{\frac{\kappa}{a_4}}, \quad t'=\frac{t}{t_0}, \quad \rr'=\frac{\rr}{l_0},\\
& W=t_0\Gamma, \quad R'=\frac{R}{l_0}, \quad \lambda=\frac{c_++c_--2c_L}{c_+-c_-}.
\end{aligned}
\right.
\end{equation}
Here, $V_{\text{at}}$ is the mean atomic volume of the alloy's species. $\lambda$ can be seen as the reduced asymmetry parameter of the 2:3:4 potential $f_h$ in \refeq{nrjhomo}. $R'$ is the atomic displacement range ratio  between ballistic relocations and diffusion. Finally, $W$ is the reduced intensity of ballistic effects. It yields the frequency of ballistic relocations $\Gamma$, and diffusion migration $M$. Prime notations are dropped in the following, except for $R'$, as both $R$ and $R'$  will be used. 

Finally, \refeq{chphi} was not the most suitable form for the elaboration of a phase field numerical scheme. For that purpose, it was rewritten under variational form:

\begin{equation}
\label{chvaria}
\left\{
\begin{aligned}
&\frac{\partial \phi}{\partial t} = \nabla^2\left[\frac{\delta \LL}{\delta \phi}(\phi)\right],\\
&\LL(\phi)\equiv F(\phi)+\frac{W}{2}\int_{\Omega}\{g*\phi\}(\rr,t)\phi(\rr,t)\dd \rr,
\end{aligned}
\right.
\end{equation}
where as $-\nabla^2 g=\delta-p_R$. The insight of \refeq{chvaria} is to interpret the dynamics of the system, as the minimization of an effective energy for the irradiated system, as $\LL(\phi)$ is a strict Lyapunov function. Stationary states can thereupon be seen as minima of $\LL$ \cite{lavrskyi2014fraton}.

%%%%%%%%%%%%%%%%
\section{Simulation Method}
\label{sec:simu}

%We present the numerical methods at use for parameters determination, and phase field simulations. 

\subsection{Parameters simulation}
\label{subsec:para}

\subsubsection{Equilibrium parameters}
\label{subsubsec:eq}

Coefficients $a_2$, $a_3$ and $a_4$, defining the free energy density $f_h(c)$ in \refeq{nrjhomo}, were estimated using the experimental AgCu phase diagram from \cite{murray1984}. Indeed, \refeq{solub} links these parameters to the experimental solubility limits  $c_{\pm}$, for a given temperature T.  Yet, \refeq{solub} only provides with two relations, which is insufficient to compute three parameters. A third constraint should thus be set. This was given by an additional assumption on the factorization of $f_h(c)$.  By construction, $f_h(c)$ is divided in a symmetric part $\omega(c)$, this is the grand-canonical potential density, and an asymmetric contribution $\Delta \mu c/V_{\text{at}}$, corresponding to the chemical potential gap between atomic species. In the case of AgCu, $\Delta \mu=-0.44$ eV  between silver and copper \cite{briki2013}, and $V_{\text{at}}=13.4$ \AA{}$^3$. The factorization assumption was  made on $\omega(c)$, namely $\omega(c)=D(c-c_+)^2(c-c_-)^2$. This way, $\omega(c)$ was minimized by the solubility limits satisfying $\omega'(c_+)=\omega'(c_-)=0$, and both wells of $\omega(c)$ had the same depth $\omega(c_+)=\omega(c_-)$. This provided with $f_h(c)=D(c-c_+)^2(c-c_-)^2+\Delta \mu c/V_{\text{at}}$, and consequently $a_2$, $a_3$ and $a_4$.

The interface profile and the stiffness parameter $\kappa$ were determined using a mixed molecular dynamics-Monte Carlo approach, for $T=600$, 700, 800 and 900 K. The interatomic potential computed by Briki \cite{briki2013} was chosen, as it led to a correct phase diagram for AgCu. The simulation sample was made of a silver and a copper block, stuck in  the $<$100$>$ direction, for a total of 11700 atoms. The tangential constraint due to the misfit between Ag and Cu cell parameter ($\sim 13$ \% \cite{shao2013}) was geometrically minimized with a $8/9$ ratio for the number of atomic rows between silver and copper. The relaxation of the remaining tangential constraint was performed using a MD simulation at 0 K, using the Verlet algorithm with time step $\Delta t=10^{-15}$ s. Computations were speeded up with the simulated annealing method. Starting from this state, the equilibrium interdiffusion concentration profile of silver at the interface between blocks was determined using a Monte Carlo method in the pseudo grand canonical ensemble. The displacement proposal in the Markov chain included atomic species switch to mimic interdiffusion at the interface,  as well as thermal and elastic relaxation. The acceptation-reject probability was computed using the Gibbs potential derived from the same interatomic potential as in the previous step, and $\Delta \mu=0.44$ eV \cite{briki2013}. Afterwards, these profiles were fitted with a Jacobi elliptic sine function \cite{Saxena1991}. The fit was performed with respect to both $\kappa$ and the form coefficient of the elliptic sine, governing the nature of the interface. It eventually provided with $\kappa(T)$ (see appendix \ref{appendix:kappa}). 

It should be noted that $\kappa$ was determined in the $<$100$>$ direction only, and $<$110$>$ and $<$111$>$ directions were passed over. The absence of the $<$110$>$ direction can be easily justified by the fact it is not an equilibrium surface of the crystal, contrary to $<$100$>$ and $<$111$>$ directions. Besides, the $<$111$>$ surface displays the smallest energy, and $<$100$>$ could hence be considered as the surface representing for high energy interfaces, when no elastic contribution is introduced.

\subsubsection{Displacement cascade simulation}
\label{subsubsec:cascade}

Displacement cascades were simulated using the Binary Collision Approximation code MARLOWE. This provided with the ballistic relocation frequency $\Gamma$, the mean relocation range $R$, and the irradiation enhanced mobility $M$. Contrary to MD simulations, MARLOWE is capable of simulating high energy incident ions irradiation required to achieve microstructure patterning \cite{Krasnochtchekov2005,Chee2010}. Besides, it accounts for crystal structure and associated atomic effects, such as channeling and replacement sequences. This explains why MARLOWE's simulations often match molecular dynamics \cite{demange2016}, yet requiring less computation time \cite{mdbcacomp}. The nuclear stopping power was the standard Moli\'ere potential, and the electronic stopping potential was taken from SRIM semi-empirical inelastic potential \cite{docsrim}. Simulations were performed on homogeneous polycrystalline Ag$_{50}$Cu$_{50}$ subjected to 1 MeV Krypton incident ions in orthogonal incidence.

\subsection{Phase field simulations}
\label{subsec:pf}

Phase field simulations on \refeq{chtot} were performed in two dimensions using a finite difference scheme in time, coupled with a spectral treatment of space. The phase field numerical scheme was elaborated to solve \refeq{chphi} for long simulation times, where potential stationary states are reached. \refeq{chphi} involves fourth order space derivatives. This leads to an overcontraining stability condition on the time step $\Delta t$, if treated explicitly. For that purpose, we chose to adapt Eyre's semi implicit  time scheme  \cite{Eyre1998,Vollmayr-Lee2003}, for time dependent gradient systems \cite{stuart1994} to our equation. It profits with the possibility to write \refeq{chphi} under variational form in \refeq{chvaria}. Following Eyre's algorithm, $\LL(\phi)$ was split into a contracting and an expansive part. The first one was treated implicitly in order to allow the time scheme to inherit with this flow-contracting  property, the second one was treated explicitly. The splitting between explicit and implicit term was  linear, so that the time scheme was compatible with a spectral treatment of space. The resulting scheme was unconditionally gradient stable, as it inherited \cite{stuart1994} with the global Lyapunov stability of the continuous solutions \cite{Song2009}. This choice led to a first order in time. 

Besides, we chose the Fourier collocation treatment of space \cite{Ye2002}. This choice allowed to get rid of spatial derivatives, thus achieving faster computation, and lead to an infinite convergence order in space. The final scheme for each $\vect{k}$ mode of the discrete solution in Fourier space after $n$ iterations  $(\hat{\phi})_{\vect{k}}^n$ reads:

\begin{equation}
\label{splitspect2}
\hat{\phi}_{\vect{k}}^{n+1}=\frac{1}{1-\Delta t\rho(\vect{k})}\left(\hat{\phi}_{\vect{k}}^n+\Delta t  \left\{a_{\vect{k}}\left[\hat{\phi}_{\vect{k}}^n\right]-\rho(\vect{k})\hat{\phi}_{\vect{k}}^n\right\}\right),
\end{equation} 
where $\rho(\vect{k})\equiv-A_c |\vect{k}|^2-|\vect{k}|^4- W (1-\hat{p}_R(\vect{k}))$ and $a_{\vect{k}}\left[\hat{\phi}_{\vect{k}}^n\right]$ $=-|\vect{k}|^2\left((3\lambda^2-1)\hat{\phi}_{\vect{k}}^n-3\lambda [\hat{\phi}_{\vect{k}}^n]^{*2}+[\hat{\phi}_{\vect{k}}^n]^{*3}+|\vect{k}|^2\hat{\phi}_{\vect{k}}^n\right)-W(1-\hat{p}_R(\vect{k}))\hat{\phi}_{\vect{k}}^n$.
  
Computations were realized on a 1000 $\times$ 1000 simulation grid. This way, at least three points were used to describe interfaces for a corresponding  100 nm $\times$ 100 nm simulation box in real units at room temperature. The numerical scheme is unconditionally stable. The only criterion for $\Delta t$ was hence the accuracy of the program for long time simulations. We chose $\Delta t=0.1$ in reduced coordinates ($\sim 0.01$ s at room temperature). 

Simulations were performed using aleatory initial conditions.

%%%%%%%%%%%%%%%%%%%%%%%%%%%%%%%%%%%%%%%%%%%%%%%%%%%%%%%%%%%%%%%%%%%%%%%%%%%%%%%%%%%%%%%%%%%%%%%%%%%%%%%%%%%%%%%%%%%

\section{Results}
\label{sec:results}

\subsection{Parameters computation}
\label{subsec:parametrization}

To show the potentiality of our model to give a quantitative description of the evolution of a system under irradiation, the microstructure evolution under irradiation in AgCu alloy was modelled. This material was chosen for two reasons. First, the silver copper system displays a miscibility gap for temperatures inferior to the critical temperature $T_c=1385$ K. Besides, AgCu, Ag, and Cu crystals are invariant under the same space group Fm$\bar{3}$m. As such, the silver-copper alloy undergoes an isostructural phase transition when the binodal is crossed. This system can hence be characterized by a scalar variable $c(\rr,t)=c_{\text{Ag}}(\rr,t)$, and its evolution is governed by \refeq{chtot}. Second, the strong contrast factor between copper and silver makes this alloy compatible with diffraction measures. 

\subsubsection{Equilibrium parameters}
\label{subsubsec:eq}

%%ICI
\paragraph{Free energy density $f_h$:}

\begin{figure}[ht]
\centering
\includegraphics[width=8.3cm]{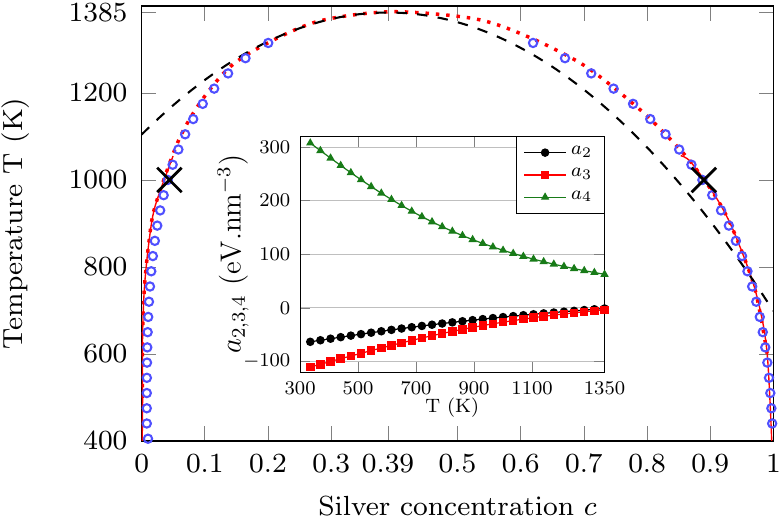}
\caption{Experimental phase diagram of silver-copper (dotted line), and comparison with the binodal curve associated to a 2:3:4 potential (present work, circles), and with a standard 2:4 symmetric potential (dashed line). Insert: Landau coefficients $a_2$, $a_3$ and $a_4$, defining the free energy density $f_h(c)$, fitted on the experimental phase diagram of silver-copper \cite{murray1984}.}
\label{fig:diagphase} 
\end{figure}

Computing the solubility limits for various temperatures provided us with the binodal curve presented in \reffig{fig:diagphase} (circle). By construction, it matches the experimental binodal curve (dotted line). This eventually provided us with $a_2(T)$, $a_3(T)$ and $a_4(T)$ (insert of \reffig{fig:diagphase}).

As required by  Landau theory, $a_4(T)$ is always strictly positive, while $a_2<0$ for $T<T_c$ and cancels for $T=T_c$.  At this temperature, the system undergoes a phase transition. For AgCu, it takes place at the critical mean concentration of silver  $\bar{c}=c_L=0.39$ \cite{murray1984}. Consistently with Landau theory, $a_2(T)$ can be approximated by a linear function close to $T_c$: $a_2(T)\simeq A(T-T_c)$ \cite{tokar2003}. Here, the best linear fitting ($\rho=0.99$) gave $A=4.8\times 10^{-2}$ ($\pm0.05$) eV.K$^{-1}$.nm{}$^{-3}$. As for the cubic term coefficient $a_3(T)$, in our case it is not negligible as the system undergoes a first order phase transition when the binodal is crossed. It only cancels at the Landau point $(c_L,T_c)$, where the phase transition becomes of second order. As displayed in \reffig{fig:diagphase}, the binodal curve associated to the 2:3:4 potential (circle) intrinsically matches the experimental binodal curve (line). In \reffig{fig:diagphase}, we also plotted the binodal curve associated to the standard Landau 2:4 potential from \cite{enrique2001compositional,Enrique2003} (dashes). Comparison with the experimental curve confirms that the cubic term in our potential is essential to account for the asymmetry of the experimental binodal curve. Indeed, it can be noted that the right branch of the symmetric 2:4 potential displays a strong discrepancy, even for high temperatures. 

\paragraph{Interfaces and stiffness parameter $\kappa$:}
\label{subsubsec:kappa}

\begin{figure}[ht]
\centering
\includegraphics[width=8.5cm]{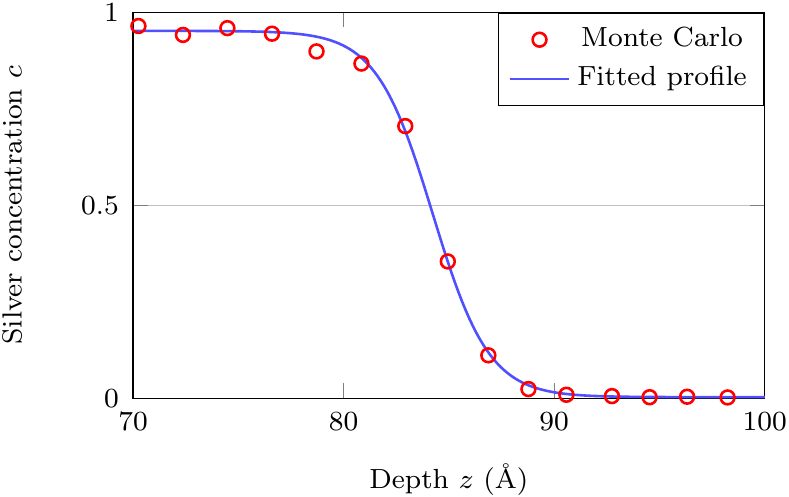}
\caption{Silver concentration profile at the interface, determined by our Monte Carlo simulation (circle) for $T=700$ K, and corresponding phase field profile (line) based on a fit by a Jacobi elliptic sine.}
\label{fig:mc} 
\end{figure}

The form coefficient of the fitted interface was found to be extremely close to 1, corresponding to the hyperbolic tangent profile. Our fitted interface profile thus intrinsically matched the physical interface given by MC simulations. This correspondance of the fitted (line), and the MC (circles) profiles is illustrated in \reffig{fig:mc},  at $T=700$ K.

The interface width and the value of $\kappa$ for different temperatures are gathered in table \ref{table:kappa}. It can be seen that the interface width increases with temperature: from $e=5.64$ \AA{} at $T=600$ K, to $e=7.68$ \AA{} at $T=900$ K. This range of width is consistent with previous simulations \cite{Wynblatt1990}. Besides, the variations of $\kappa$ remain within the precision uncertainty of our MC simulations, so that $\kappa$ can be considered constant. From an average over every MC simulation, we could set $\kappa=0.15$ ($\pm 0.005$) eV\AA{}$^{-1}$. This result seems to deviate from the standard Landau-Ginzburg formalism, where $\kappa$ is supposed to decrease to 0 as $(T-T_c)^{1/2}$, when T approaches to $T_c$. Actually, our MC simulations were performed for temperatures up to 900 K only. The corresponding constant value for $\kappa$ cannot hence be transferred above 1000 K, restraining the study to temperatures sensibly inferior to $T_c=1385$ K. This is actually not much of a restriction, considering that the silver copper alloy is in liquid phase above 1052 K.

\begin{table}[ht]
\centering
\begin{tabular}{l l l l l l l}
   \hline
 T (K) &   $e$ (\AA{}) & $\kappa$ (eV.\AA{}$^{-1}$) & $c_-$ & $c_+$ &  $\rho$ (corr.)\\
   \hline	
   \hline
600     & 5.64 $\pm 0.01$& 0.148 $\pm 0.005$ & $ 0.004$&$0.97$ $\pm 0.01$ & 0.9996 \\
700     & 6.18 $\pm 0.01$& 0.146 $\pm 0.005$ & $0.004$&$0.96$ $\pm 0.01$ & 0.9997 \\
800     & 6.84 $\pm 0.01$& 0.146 $\pm 0.005$ & $ 0.004$&$0.95$ $\pm 0.01$ & 0.9998 \\
900     & 7.68 $\pm 0.01$& 0.144 $\pm 0.005$ & $0.009$&$0.94$  $\pm 0.01$ & 0.9998 \\
   \hline

\end{tabular}

\caption{Interface width $e$ (\AA{}) and stiffness parameter $\kappa$ (eV.\AA{}$^{-1}$) from our silver concentration  phase field profile, after fitting on our Monte Carlo simulations, and solubility limits $c_{\pm}$ from our Monte Carlo simulations, for various temperatures.}
\label{table:kappa}
\end{table}

As expected, our atomistic simulations account for the asymmetry of silver solubility limits $c_{\pm}$. In particular, copper enrichment in silver phase for high temperature is patent in table \ref{table:kappa}. For example, at $T=900$ K, copper concentration in silver phase is $1-c_+=0.06$.

\subsubsection{Irradiation parameters}
\label{subsec:irr}

\paragraph{Ballistic mixing parameters $\Gamma$ and $R$:}
\label{subsubsec:bal}

%%percolation
The preliminary step before computing $\Gamma$ and $R$, was the validation of the percolation criterion \refeq{percol}. In average, we found $L_{\text{c}}= 210$ nm for our cascades, within which $n_{\text{sc}}= 6$ subcascades of mean size $l_{\text{sc}}\geq 15$ nm were observed. This is consistent with primary knocked atoms (PKA) energies ranging from 80 keV to 180 keV in our simulations \cite{Metelkin1997}. This led to a covering fraction $f_r= 0.4$, thus sufficient to ensure the covering.

\begin{figure}[ht]
\centering
\includegraphics[width=8cm]{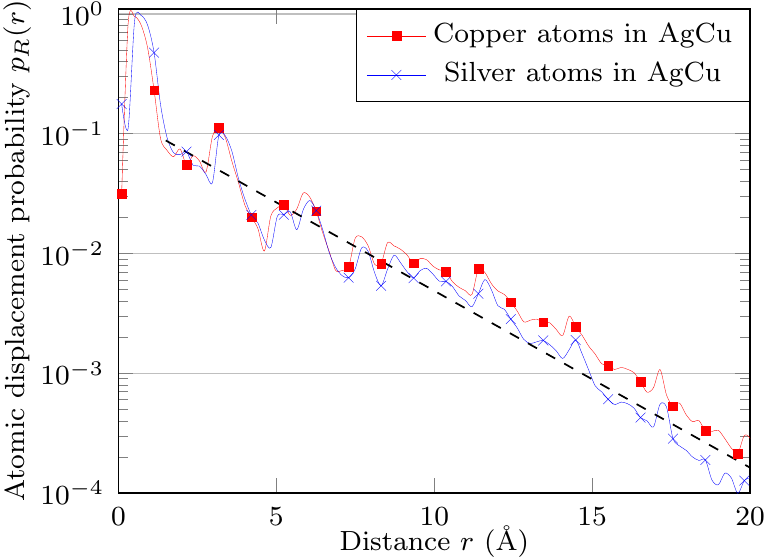}
\caption{Normalized distribution of atomic relocation of copper and silver atoms, in polycrystalline Ag$_{50}$Cu$_{50}$ irradiated by Krypton ions of energy 1 MeV, obtained with the BCA code MARLOWE. Average over 1000 trials. Silver and copper distributions display almost equivalent distributions, fitted ($\rho=0.97$) by a unique exponential distribution of slope $1/R$, $R=3.04$ \AA{} (dashes).}
\label{fig:dep} 
\end{figure} 

It was hence meaningful to display in \reffig{fig:dep} the normalized distribution of displacements defined by \refeq{loidep1}, for copper (cross) and silver (square) atoms. Obviously, the exponential decay assumption (see \refeq{loidep1}) is validated ($\rho=0.97$) for both species. Note however the periodically distributed peaks in both distributions. These are actually due to a lower binding energy barrier required for an atom to migrate to a neighbouring site.  Besides, the silver and copper atoms distributions are very close. This allowed us to use  $R=3.04$ \AA{} for both species (dashes). This common value of $R$ is dependent on incident ions species, but not on their initial energy. In our phase field simulations, we hence used $R=3.04$ \AA{} for every flux and temperature. This value matches a previous mixed  MDrange-molecular dynamics simulation \cite{Enrique2003} where $R=3.08$ \AA{}.

The simulations also provided with the total amount of relocated atoms per cascade $N_d=309000$. Using this result with  $L_{\text{c}}= 210$ nm, we could compute the relocation cross section $\sigma^d=2.01\times 10^{-13}$ cm$^2$. An experimental element of comparison can be found in  \cite{bellon2001}. For an incident flux $\Phi=1.21\times 10^{13}$ cm$^{-2}$s$^{-1}$ of 1 MeV Krypton ions on polycrystalline AgCu, Enrique and Bellon could measure $\Gamma=2.8$ s$^{-1}$. Using the same flux, our model shows excellent consistency: $\Gamma=\sigma^d \Phi=2.53$ s$^{-1}$.

\paragraph{Atomic mobility $M$ under irradiation:}
\label{subsubsec:dif}

\begin{figure}[ht]
\centering
\subfigure[~~~Atomic mobility]{\includegraphics[width=4.2cm]{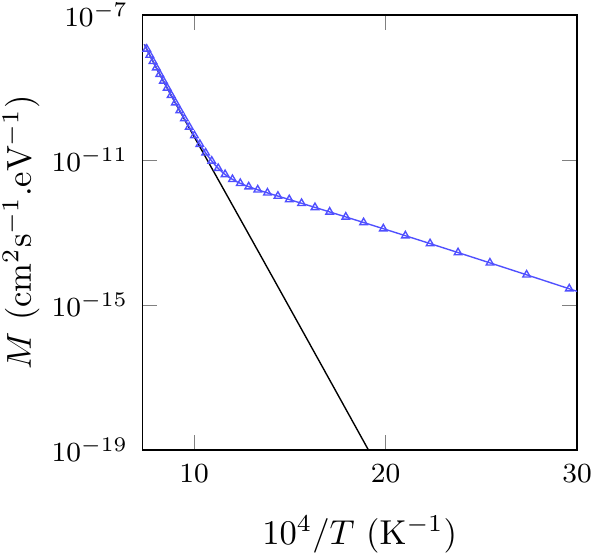}\label{fig:mob}}
\subfigure[~~~Time scale]{\includegraphics[width=4.2cm]{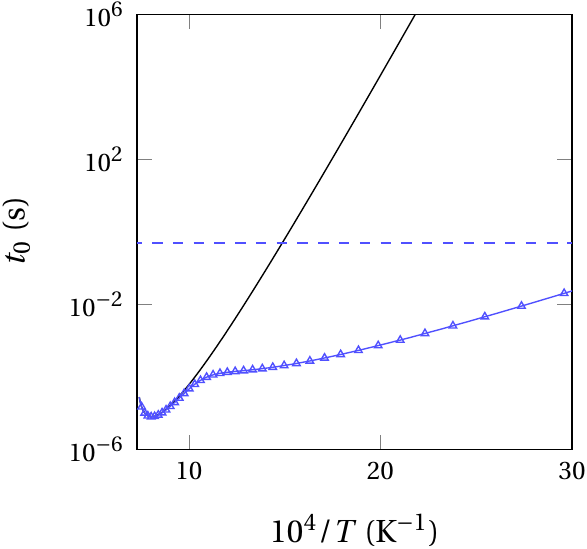}\label{fig:t0} }
\caption{Atomic mobility (left) and characteristic time scale (right) of silver and copper atoms in AgCu, without irradiation (line), and for $\Phi= 10^{13}$ cm$^{-2}$s$^{-1}$ (triangle), and time scale of ballistic exchange $\Gamma^{-1}$ for $\Phi= 10^{13}$ cm$^{-2}$s$^{-1}$ (dashed line) }
\label{fig:mobtot} 
\end{figure}

From MARLOWE simulations for 1 MeV Krypton ions, we also extracted the number of Frenkel pairs per cascade before thermal annihilation  and sink absorption $N_V=19500$. Injecting diffusion numerical data \cite{Bandyopadhyay1990,Foiles1986,Nordlund1998,Adams1988,Wynblatt1969,Mcgervey1973,McGervey1975} in \refeq{mobth}, \refeq{mobirr} and \refeq{coefdiffedcv3}, we could compute the total mobility for any flux and temperatures. \reffig{fig:mob} compares the atomic mobility without irradiation (line), and for a flux $\Phi= 10^{13}$ cm$^{-2}$s$^{-1}$ (triangles). 
 
The amplitude of $M$ under irradiation matches the thermal mobility for high temperatures ($T>900$ K). Indeed, both follow the same Arrhenius law, and no diffusion enhancement is observed. It can be explained by the saturation of thermal point defects. Yet, a kinking arises when T decreases below 900 K, as the mobility under irradiation becomes by several orders superior to the equilibrium mobility. This observation is of particular matter, as the mobility plays a central part in the characteristic relaxation time of diffusion $t_0$ (see \refeq{adim1}). As such, \reffig{fig:t0} compares $t_0$ to the characteristic time scale of ballistic relocations in the cascade, $\Gamma^{-1}=0.498$ s$^{-1}$ (for $\Phi= 10^{13}$ cm$^{-2}$s$^{-1}$). This allows to investigate the range of temperatures where $t_0$ and $\Gamma^{-1}$ are close, so that competition between diffusion effects and ballistic relocations is possible. In the case of equilibrium diffusion (line), this happens above 700 K, versus around 300 K in the present irradiation enhanced mobility model (triangle). This difference will be discussed in section \ref{sec:results}.

%Note besides that a similar reasoning highlights how defects absorption at interfaces is essential. Indeed, we observed it counterbalances the dramatic rise of interstitials under irradiation, otherwise leading to an atomic diffusion overestimation for low temperatures.

\subsection{Phase field simulations in reduced coordinates}
\label{subsec:simu}

\begin{figure}[ht]
\centering
\subfigure[~~Phase separation $\Gamma\ll 1/t_0$]{\includegraphics[width=4cm]{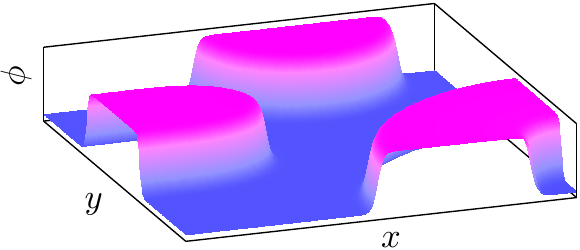}\label{fig:3d3}}
\subfigure[~~Patterning $\Gamma\sim 1/t_0$]{\includegraphics[width=4cm]{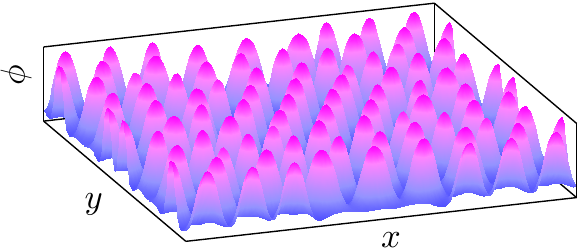}\label{fig:3d2}}
\subfigure[~~Solid solution $\Gamma\gg 1/t_0$]{\includegraphics[width=4cm]{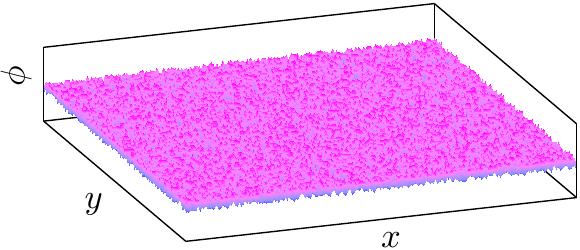}\label{fig:3d1}}
%\subfigure[~~Precipitates radius]{\includegraphics[width=4cm]{fig12}\label{fig:3d4}}
\caption{Reduced concentration $\phi$ from our phase field simulations: aleatory initial condition with $\bar{\phi}=0$ ($c(t=0)=c_L$) and $\lambda=0.25$ ($T<300$ K), on a 1000 $\times$ 1000 mesh grid, with $\Delta t=0.1$ and $t_{\text{final}}=10^6$. Top left: demixtion $(W,R')=(0.005,2.4)$. Top right: patterning $(W,R')=(0.12,2.4)$. Bottom: solid solution $(W,R')=(1,2.4)$.}
\label{fig:3d} 
\end{figure}

\begin{figure}[ht]
\centering
\includegraphics[width=8cm]{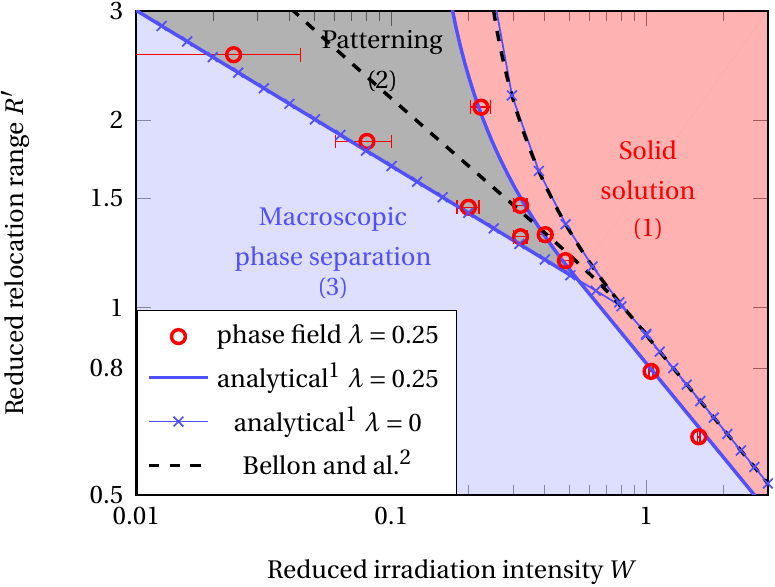}
\caption{Phase diagram in reduced coordinates, displaying three microstructure behaviours: solid solution (top right), patterning (top center) and phase separation (bottom left). Comparison between the present phase field study's numerical results (circles) and analytical results$^1$ from \cite{champ} with $\lambda=0.25$ (thick line), $\lambda=0.25$ ($\times$ mark), and Bellon and al.'s diagram from \cite{bellontheo} (dashed line).}
\label{fig:phase} 
\end{figure}

By the direct observation of the copper precipitates that appeared in our simulations (\reffig{fig:3d3}, \reffig{fig:3d2} and \reffig{fig:3d1}) for long times ($t=10^6$ in reduced time), three evolutions of the microstructure could be distinguished. Each microstructure  defines a  domain in the $(W,R')$ plane in \reffig{fig:phase}. $\lambda=0.25$ and $R'=2.4$ were set for the 3 microstructure examples displayed in \reffig{fig:3d}, as they correspond to room temperature in real units, in  the case of AgCu under 1 MeV Krypton ions.

For high values of $W$ (\reffig{fig:3d1}, domain (1) in \reffig{fig:phase}), we observed a homogeneous system corresponding to the formation of a solid solution. In this case, the disordering due to long range displacements induced by irradiation overcomes the expected phase separation driven by  thermal diffusion: $\Gamma\gg 1/t_0$. This corresponds to a precipitates radius close to 0 at all time. On the contrary, for small values of $W$ (\reffig{fig:3d3}, domain (3)  in \reffig{fig:phase}), diffusion ordering effects are predominant as $\Gamma\ll 1/t_0$. The system displays a macroscopic phase separation, similar to Ostwald ripening \cite{Lifshitz1961}. This corresponds to an unbounded growth of precipitates radius. Between these two domains lies the patterning domain (2) (\reffig{fig:3d2}). In this domain, precipitates reach a finite size and remain stationary. This is due to ballistic and diffusional effects coming to equilibrate one another. It is associated to a  mean precipitates radius quickly reaching its finite asymptotic value. 

The numerical limits between these three domains (circles in \reffig{fig:phase}) were determined using three criteria. The first one was the direct observation of solutions. The second one was the time evolution of precipitates mean size on a sample of 200 solutions from random initial conditions. The last criterion was the evolution  of the Lyapunov function $\LL(\phi)$ (see \refeq{chvaria}). Indeed, $\LL(\phi)$ converges to a non null limit in the patterning domain, but keeps decreasing in case of macroscopic phase separation, albeit slowly. Numerical limits were superimposed on the three analytical ones from a self-consistent approach \cite{champ} (continuous line).  Both methods perfectly match. The advantage of our numerical phase field study, is that contrary to the self-consistent analytical work \cite{champ}, it does not require further simplification such as linearization of \refeq{chphi}. It should be noted that contrary to chemical disordering, ballistic disordering is of finite range $R$, and patterning can only be achieved for a sufficient value of $R$. Indeed, it allows ballistic relocations to display longer range than diffusion migrations ($R\gg l_0$). We also plotted Bellon and al. domain limits based on a spectral method \cite{bellontheo} (dashed lines). Though qualitatively similar, the two diagrams quantitatively diverge. This disparity with our results has two origins. First, Bellon's limit (dashed line) between solid solution (1) and domain formation (2)+(3) differs from ours for $\lambda=0.25$ (thick line) by a translation operation, due to the use of a symmetric 2:4 potential in \cite{bellontheo} ($\lambda=0$ in reduced coordinates). It should be noted that, in the case of AgCu, depending on the temperature, $\lambda$ varies between 0.15 and 0.25, and it cannot hence be set to 0.  This discrepancy emphasizes the necessity for an asymmetric potential based on the experimental phase diagram of AgCu. Second, Bellon used a mixed tangent-sine ansatz to study the crossing of the demixtion (3)-patterning (2) limit, eventually leading to a smaller patterning domain. Actually, by restraining the concentration profile to particular ansatz in  their prospection for patterned microstructure \cite{bellontheo}, they missed a variety of patterned microstructure. These mircostructures correspond to the phase diagram area between the dashed line and the tick line.   This can be seen as this limit differs from our estimation at $\lambda=0$ ($\times$ marks).

\subsection{Phase field simulations in real units: phase diagram under irradiation in $(T,\Phi)$ plane}
\label{subsec:simu2}

\begin{figure}[ht]
\centering
\includegraphics[width=8cm]{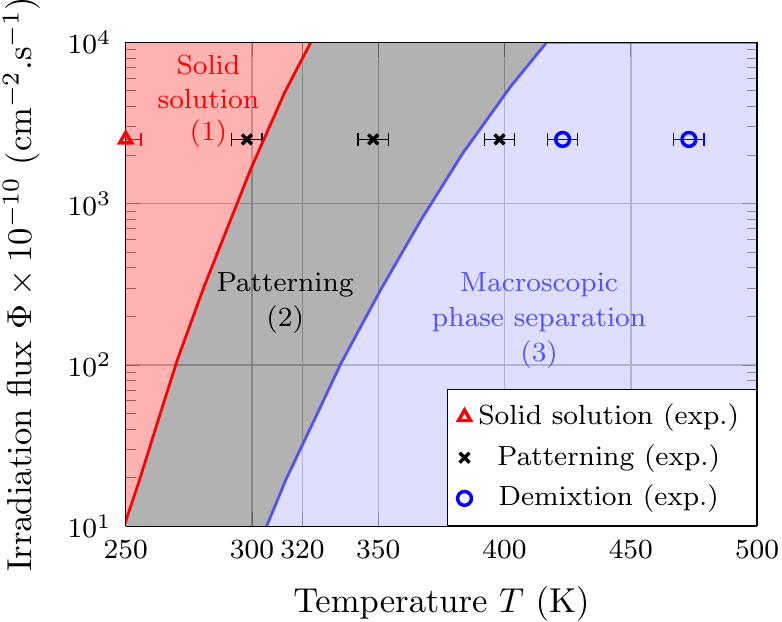}
\caption{Phase diagram in the temperature-flux plane for the silver-copper system subjected to Krypton ions (1 MeV) irradiation, displaying three microstructure behaviours: solid solution (top left), patterning (top center) and phase separation (right). For low temperatures and moderate flux, the system displays stationary precipitates (disrupted coarsening). Circle, triangle and cross: diffraction results from Wei and al. \cite{Wei1997}}
\label{fig:phaseagcu} 
\end{figure}

In the case of AgCu irradiated by Krypton 1 MeV incident ions, the reduced diagram in \reffig{fig:phase} can be transposed in real units. For that purpose, we used that $R=3.04$ \AA{} regardless of $\Phi$ and $T$. As for $W$, it is fully determined by $\Phi$ and $T$, through $M(\Phi,T)$ (\reffig{fig:mob}), $a_{2,3,4}(T)$ (insert \reffig{fig:phase}), $\Gamma(\Phi)$ and $\kappa$. This provided us with the phase diagram of silver-copper under irradiation, in the temperature-flux plane displayed in \reffig{fig:phaseagcu}. This result is of great interest, as this diagram can be directly compared to experiments. To our knowledge, no quantitative and predictive study of disrupted coarsening under irradiation  had been performed so far, at least in real units.

The three microstructural domains displayed in \reffig{fig:phase} are present. The formation of a solid solution of AgCu is achieved for high fluxed and low temperatures (1) where ballistic disordering overcomes diffusion. Complete phase separation (3) occurs for high temperature and moderate fluxes. In these conditions, diffusion ordering effects are predominant. Finally, the  patterning domain (2) requires moderate flux and temperature. This allows equilibrium between ion mixing and diffusion. Interestingly, patterning effects can be achieved at room temperature, for realistic fluxes of 1 MeV Krypton ions. This made experimental validation possible. 

%We also plotted the diagram extrapolated from Bellon and al. \cite{bellontheo}. , using their  reduced irradiation phase diagram in  $(R,W)$ plane , their atomic mobility and potential \cite{enrique2001compositional}, their experimental relocation cross section  \cite{bellon2001} and numerical estimation of the mean relocation range in subcascades \cite{Enrique2003}, led to a 300 K right shift of patterning domain limits, as well as an extremely thin patterning domain. This discrepancy could be alleviated with our model of mobility, but still remained very far from experimental data presented in the next paragraph. To a large extent, this is due to the accuracy loss of the standard 2:4 Landau potential for low temperatures.  

Wei and Averback \cite{Wei1997} performed diffraction experiments on the silver-copper system, irradiated with a $\Phi=2.5\times 10^{13}$ cm$^{-2}$s$^{-1}$ flux of 1 MeV Krypton ions. Their prospecting of eventual steady states were performed for a maximal final dose of $1.9\times 10^{16}$ cm$^{-2}$. \reffig{fig:phaseagcu} displays excellent consistency between our numerical results (continuous line) and their experiments (cross, triangle and circle). 

In details, they observed the formation of a solid solution for $T=80$ K (triangle), complete separation of phases for $T=473$ K (circle), and patterning for $T=298,348$ and $398$ K (cross). Among these points, the solid solution for $T=80$ K (liquid nitrogen), the complete separation of phases for $T=473$ K and the patterning for $T=348$ K, perfectly match our diagram. Besides the two other patterning points ($T=298$ K and $T=398$ K) remain within the uncertainty, notably coming from the choice of migration energies in the atomic mobility in our simulations, as well as the temperature standard deviation of diffraction measures. Another point corresponding to phase separation is displayed on the diagram:  for $T=423$ K, Wei and Averback observed an intermediate microstructure between a patterning ($T=398$ K) and phase separation ($T=473$ K). They could not conclude whether it was a true steady state or a slow phase separation, as asserted by  Bellon \cite{bellontheo}.

We actually see two reasons why Wei and al. failed to decide whether $T=423$ K lead to patterning or phase separation. The first one is of experimental nature as discussed in \cite{Wei1997}. For this temperature, they could not afford greater dose than $1.9\times 10^{16}$ cm$^{-2}$: such doses would lead to experimental artifacts, such as pure copper phase induced by sputtering. Considering this dose was achieved with $\Phi=2.5\times 10^{13}$ cm$^{-2}$s$^{-1}$, the system was followed during 26 minutes top. Still, our study estimated that the characteristic time scale for the microstructure evolution for $T=423$ K and $\Phi=5\times10^{12}$ cm$^{-2}$s$^{-1}$ was $t_0\sim 3\times 10^{-2}$ s. Besides, our simulations required at least $10^6$ iterations with $\Delta t=0.1$ ($\times t_0$), viz almost an hour in real time for the system to reach its final state. 26 minutes was hence barely sufficient to reach the final state in case of phase separation. Moreover, in our study, steady state was always reached within $10^5$ iterations in case of patterning. This corresponds to 5 minutes at this temperature. And so, 26 minutes should have been sufficient for the system to reach its steady state otherwise. For that reason, accordingly with Bellon, this flux-temperature couple leads to phase separation indeed. Second, the limit between patterning and demixing domains is continuous, contrary to the the solid solution-patterning limit: when the limit between the patterning and the demixing domain is crossed,  solution profiles vary continuously. This limit hence rather behaves like an overlapping. This suggests the limit is intrinsically blurred. This makes the distinction between complete phase separation and patterning quite uneasy, whether it is prospected numerically or experimentally.

\section{Conclusions}
\label{sec:dis}

In this work, the influence of irradiation on steady state patterned  microstructures in binary alloys was studied, using a multiscale approach based on phase field. The methodology was applied to the silver-copper system, under Krypton ions irradiation.

First, a phase field model adding an athermal irradiation term \cite{martin1984} based on the ion mixing formalism \cite{sigmund1981} was set. Contrary to most phase field studies, this model could account for a cubic term in the free energy density. 
Then, an innovative finite difference scheme \cite{Eyre1998} was developed, allowing to simulate the late stage microstructure of an irradiated binary alloy in reduced units. Besides, atomistic methods were implemented to compute the phenomenological parameters in the case of AgCu irradiated with 1 MeV Krypton ions. Phase field simulations could hence be transferred to real space and time units. 

As a result, the temperature and flux leading to patterning in AgCu subjected to 1 MeV Krypton ions was determined. This was summarized in a phase diagram in the flux-temperature plane, which highlighted a patterned microstructure domain, for temperature ranging from 250 K to 400 K, and incident fluxes greater than $10^{10}$ cm$^{-2}$s$^{-1}$. This diagram was in excellent agreement with diffraction experiments \cite{Wei1997}. To our knowledge, no previous numerical study could achieve this diagram for an irradiated material.

\bibliographystyle{apsrev4-1} 
\bibliography{bibliographie}

\appendix

\section{Expression of $\kappa$}
\label{appendix:kappa}

Without irradiation, the interface phase field profile is given by The Jacobi elliptic sine \cite{demangethese2015}:

\begin{equation}
\label{soljacob1}
\eta^{(p)}(x)=A^{(p)}\jaco\left[\frac{x}{\Delta^{(p)}},p\right]-\frac{a_3}{3a_4},
\end{equation}
where $p\in ]0,1]$ is the form factor. The amplitude $A^{(p)}$ and the half width $\Delta^{(p)}$ are given by:

\begin{equation}
\label{epparoiasym1}
A^{(p)}=\alpha\sqrt{\frac{2p^2}{1+p^2}}, \quad \Delta^{(p)}=\frac{1}{\sqrt{1+\frac{a_3^2}{3a_4 |a_2|}}} \sqrt{1+p^2}\sqrt{\frac{\kappa}{a_2}}.
\end{equation}
For long times, $p\to 1$ and $\jaco(\cdot,p)$ becomes an hyperbolic tangent, so that the interface width $e=2\Delta^{(1)}$ reads: 

\begin{equation}
\label{epparoiasym1}
e=\frac{1}{\sqrt{1+\frac{a_3^2}{3a_4 |a_2|}}} 2\sqrt{2}\sqrt{\frac{\kappa}{a_2}}.
\end{equation}

\section{Expression of the mobility}
\label{ap:mob}

Without irradiation, Onsager's irreversible process formalism provides with:

\begin{equation}
\label{mobeta}
\begin{aligned}
M(\eta)&=\frac{1-\left[2(\eta+c_L)-1\right]^2}{8 k_B T}[\left(1-\left[2(\eta+c_L )-1\right]\right)D_{\text{Cu}}^{*}\\
&+\left(1+\left[2(\eta+c_L )-1\right]\right)D_{\text{Ag}}^{*}].
\end{aligned}
\end{equation}
Under irradiation, $M=\bar{M}+M^{\text{irr}}$. Based on similar diffusion coefficients of silver, copper and silver-copper alloy, and using the fact that for moderate temperatures, mainly vacancies are present at equilibrium $M^{\text{th}}$ reads:

\begin{equation}
\label{mob2ter}
M^{\text{th}}\simeq \frac{\bar{c}(1-\bar{c})}{k_B T} D_{\text{Cu}\rightarrow\text{Cu}}^{*}\simeq \frac{\bar{c}(1-\bar{c})}{k_B T} c_V^{\text{eq}} D_{V}(T),
\end{equation}
where $D_{V}(T)$ is the diffusion coefficient of vacancies in copper. This gives:
 
\begin{equation}
\label{coefdiffedcv3}
M^{\text{th}}(T)\simeq \frac{\bar{c}(1-\bar{c})}{k_B T} D_0^{V}\exp\left[-\frac{H_{V}^m+H_{V}^f-T(S_{V}^m+S_{V}^f)}{k_BT}\right].
\end{equation}
where $c_V^{\text{eq}}$ is the concentration of vacancies at equilibrium. $H_{V}^m$, $S_{V}^m$, $H_{V}^f$, and $S_{V}^f$ are the enthalpy and entropy of migration and formation of vacancies in copper respectively. As regards $M^{\text{irr}}$, it originates from  the vacancies and interstitials $\delta c_V^{\text{irr}}$ and $\delta c_{I}^{\text{irr}}$ created in the displacement cascade:

\begin{equation}
\label{cinetiquefrenkel1}
\left\{
\begin{aligned}
\frac{\dd \delta c_{I}^{\text{irr}}}{\dd t}&=\sigma^{V} \Phi-\frac{4\pi R_{\text{cap}}}{V_{\text{at}}}D_I \delta c_{I}^{\text{irr}}\delta c_{V}^{\text{irr}}-k^2D_I\delta c_{I}^{\text{irr}}\\
\frac{\dd \delta c_{V}^{\text{irr}}}{\dd t}&=\sigma^{V} \Phi-\frac{4\pi R_{\text{cap}}}{V_{\text{at}}}D_I \delta c_{I}^{\text{irr}}\delta c_{V}^{\text{irr}}-k^2D_V\delta c_{V}^{\text{irr}}.
\end{aligned}
\right.
\end{equation}
$\sigma^{V}\Phi$ is the defects rate production where $\sigma^V=V_{\text{cell}}/N_{\text{cell}}\times N_V/ L_{\text{c}}$, $4\pi R_{\text{cap}}/V_{\text{at}}D_I$ is the thermal annihilation rate of point defects, where $R_{\text{cap}}\simeq 2.5a_0$ is the recombination radius of Frenkel pairs in AgCu.  $-k^2D_{V,I}\delta c_{V,I}^{\text{irr}}$ is the sink absorption at precipitates interface, where $1/k$ is the average size of precipitates. Stationary concentration are thus given by:

\begin{equation}
\label{equilibrefrenkel1}
\delta c_I^*=\sqrt{\frac{\sigma^{V} \Phi V_{\text{at}}}{4\pi R_{\text{cap}}D_I}} \sqrt{\frac{D_V}{D_I}}, \quad \delta c_V^*=\sqrt{\frac{\sigma^{V} \Phi V_{\text{at}}}{4\pi R_{\text{cap}}D_I}} \sqrt{\frac{D_I}{D_V}}.
\end{equation}
Note the inverse size of precipitates $k$ disappears, which exonerates us from knowing the microstructure, to compute $\delta c_V^{\text{irr}}$ and $\delta c_{I}^{\text{irr}}$. We eventually get:

\begin{equation}
\label{coefdiffedcv3}
M^{\text{irr}}=\frac{2\bar{c}(1-\bar{c})}{k_B T} \sqrt{\frac{D_0^V\sigma^V V_{\text{at}}\Phi}{4\pi R_{\text{cap}}}}\exp\left[-\frac{H_{V}^m-TS_{V}^m}{2k_BT}\right].
\end{equation}

\end{document}